
\input harvmac

\Title{UCSBTH-92-14}{Gauge Symmetries and Amplitudes in N=2 Strings}
\centerline{Miao Li \foot{Email address: li@denali.physics.ucsb.edu}}
\centerline{\it Department of Physics}
\centerline{\it University of California}
\centerline{\it Santa Barbara, CA 93106}

\vskip .3in
Picture changed operators are discussed in $N=2$ strings with space-time
signature $(2,2)$. A gauge symmetry algebra is derived in a background
of torus space-time $T^{2,2}$ and its simple representation on the
picture changed operators is given. Simple Ward identities associated with
the gauge algebra and their consequences for three and four point amplitudes
of arbitrary loops are also discussed.

\Date{April, 1992}

\newsec{Introduction}

The $N=2$ string recently studied in depth in \ref\ov{H. Ooguri and C.
Vafa, Nucl. Phys. B361 (1991) 469.}, just as the $d=2$ string, is an
amusing string model. They both have a relatively simple field content.
In the case of the $N=2$ string with a flat background $R^{2,2}$,
where the superscript $(2,2)$ indicates the signature of space-time,
there is only one space-time scalar \ref\ade{M. Ademollo
et al., Phys. Lett. 62B (1976) 105; M. Ademollo et al., Nucl. Phys.
B111 (1976) 77; D. Mathur and S. Mukhi, Nucl. Phys. B302 (1988) 130;
N. Ohta and S. Osabe, Phys. Rev. D39 (1989) 1641.}. This fact can be easily
seen in a direct calculation of the one loop partition function \ov.
There are twisted versions of the symmetric string. Still, in those cases,
there are only finite many space-time scalars. Unlike the $d=2$ string,
however,
$N=2$ strings have not been studied in a nontrivial background.
Possible discrete states are therefore unknown. These potential discrete
states will certainly be important to both the study of amplitudes and
the study of the unifying structure of the $N=2$ string.

It was noted in \ov\ that the model is simple enough to be integrable.
Calculation of four point tree amplitudes shows that they are vanishing.
This is related to the fact that the low energy action derived in \ov\
is indeed an exact action. There are only nontrivial three point tree
amplitudes. Despite this, there is still no explicit computation
confirming that all higher point tree amplitudes are vanishing. Our
work presented here is motivated by an observation in a recent work
\ref\give{A. Giveon and A. Shapere, preprint CLNS-92/1139,
IASSNS-HEP-92/14.} that on a compactified background $T^{2,2}$,
there are many conserved currents on the world sheet. Ward identities
associated to these currents may prove useful in calculating amplitudes.
Although a more concrete meaning is to be understood of the off-shell algebra
proposed in \give, a somewhat well-founded algebra can be used to
derive associated Ward identities. We take the following definition of
conserved charges. They are constructed as an integral along a closed curve on
the world sheet and must be BRST invariant. The charge being conserved
should not depend on the curve, therefore the current is conserved
up to a BRST commutator. To set the framework, we shall study the
picture changed operators in section 2. This framework may not seem
necessary for our consideration of Ward identities, but it should
be useful for study of strings propagating in a general background.
It also helps one to avoid any conceptual confusions in deriving
Ward identities. The natural operators of interest are those
in picture $(-1,-1)$, just as the natural operators of the NS sector
in the $N=1$ string are those in picture $-1$.

We then, in section 3, briefly discuss an algebra of conserved charges
in a torus background $T^{2,2}$. This algebra is just the gauge algebra
of the model in this background. This being so, it follows that there are
no other physical states in addition to the known ones. This is
still a conjecture but appears very plausible, because the algebra derived
this way is closed. We shall not explore a geometric realization of this
algebra. The representation of this algebra on the picture $(-1,-1)$
operators is simple.
We proceed, in section 4, to discuss a general definition of amplitudes.
Though this definition is a bit complicated, while derivation of
Ward identities associated with symmetries discussed in section 3 is
straightforward based on this definition. The simple Ward identities
are then applied to a discussion of three point amplitudes and four
point amplitudes of arbitrary loops. Since the structure of Ward
identities is independent of the number of loops, any result obtained
from these identities is valid at any loops or even nonperturbatively.
A large class of four point amplitudes are shown to be vanishing.
More results for higher point amplitudes can be obtained by use of Ward
identities, though we shall not do so in this paper.

It is straightforward to generalize our approach to asymmetric $N=2$
strings and $N=2$ heterotic strings \ref\oov{H. Ooguri and C. Vafa,
Nucl. Phys. B367 (1991) 83.}.

\newsec{Picture Changed Operators}

As the calculation of the genus one partition function shows, there is only
one massless scalar field in the spectrum of the un-twisted N=2 string
moving in the flat background $R^{2,2}$. This scalar represents the
deformation of the K\"ahler potential. In this section various versions
of vertex operators of the massless scalar will be worked out. The discrete
states at zero momentum that appeared recently in the literature \ref\dis{
J. Bie\'nkowska, preprint EFI 91-65, 1991.} will be seen as
certain non-singular limit of the BRST invariant operators of finite
momenta.

Following Ooguri and Vafa, we introduce two complex coordinates $(x^1, x^2)$
for $R^{2,2}$ with metric $dx^1d\bar{x}^1-dx^2d\bar{x}^2$. The superspace
representation is often helpful in doing concrete calculation. Introduce
holomorphic world sheet supercoordinates $Z=(z, \theta^+, \theta^-)$ and
their anti-holomorphic partners. The left moving superfields are
\eqn\super{\eqalign{&X^i_L(z,\theta^+,\theta^-)=x^i_L(z-\theta^+\theta^-)+
\psi^i_L(z)\theta^- \cr
&\overline{X}^i_L(z,\theta^+,\theta^-)=\bar{x}^i_L(z+\theta^+\theta^-)+
\theta^+\overline{\psi}^i_L(z).}}
There are four scalars and four spin $1/2$ fermions on the world sheet,
and the total central charge of the system is 6 as it must be for a $N=2$
critical
string. The propagators are $x^i_L(z)\bar{x}^j_L(w)=-(1/2)\eta^{ij}\hbox{log}
(z-w)$ and $\psi^i_L(z)\overline{\psi}^j_L(w)=\eta^{ij}1/(z-w)$. Next write
down the supercurrent with components
$$T(Z)={1\over 2}J(z)+\theta^- G^+(z)+G^-(z)\theta^+
+\theta^+\theta^- T(z)$$
as $T(Z)=(1/2)(D^-\overline{X}_L)\cdot (D^+X_L)$, where
$D^{\pm}=\partial_{\theta
^{\mp}}+\theta^{\pm}\partial_z$. In components, $J=\psi_L\overline{\psi}_L$,
$G^+=-\psi_L\partial\bar{x}_L$, $G^-=\overline{\psi}_L\partial x_L$. Scalar
product will be always implicit in formulas.

As in bosonic strings, there is a pair of $(b, c)$ ghost of spin $(2,-1)$,
responsible for gauging world sheet reparametrization. There are in addition
two pairs of bosonic ghosts $(\beta^{\pm}, \gamma^{\mp})$ of spin
$(3/2, -1/2)$, introduced for two world sheet supersymmetries. Finally,
a pair of fermionic ghosts $(\tilde{b}, \tilde{c})$ of spin $(1,0)$ for
gauging $U(1)$ symmetry. These fields can be nicely organized into two
superfields:
\eqn\ghost{\eqalign{&B=\tilde{b}+\theta^+\beta^--\theta^-\beta^++\theta^+
\theta^-b\cr
&C=c+\theta^+\gamma^-+\theta^-\gamma^++\theta^+\theta^-\tilde{c}.}}
The propagators are organized in $B(Z_1)C(Z_2)=\theta^+_{12}\theta^-_{12}/
Z_{12}$, where $\theta^{\pm}_{12}=\theta^{\pm}_1-\theta^{\pm}_2$ and
$Z_{12}=z_1-z_2-(\theta^+_1\theta^-_2+\theta^-_1\theta^+_2)$. Again pretty
cumbersome generators of the N=2 superconformal algebra of the ghost
system can be included in a simple formula for the supercurrent, $T_g(Z)=
-(1/2)D^-BD^+C-(1/2)D^+BD^-C+\partial(CB)$.

The holomorphic part of the super vertex operator of the massless scalar
is $V_p(Z)=\hbox{exp}(i\bar{p}X_L(Z)+ip\overline{X}_L(Z))$ with on-shell
condition $p\bar{p}=0$. Written in components, it reads
\eqn\vertex{\eqalign{&V_p(Z)=\tilde{V}_p(z)+V_p^-(z)\theta^++\theta^-V_p^+(z)
+\theta^+\theta^-V_p(z)\cr
&\tilde{V}_p=e^{i\bar{p}x_L+ip\bar{x}_L}, \quad V^-_p=-i(p\overline{\psi}_L)
\tilde{V}_p\cr
&V^+_p=-i(\bar{p}\psi_L)\tilde{V}_p,\quad V_p=(ip\partial\bar{x}_L-i\bar{p}
\partial x_L+(\bar{p}\psi_L)(p\overline{\psi}_L))\tilde{V}_p.}}
The top component $V_p$ is of conformal dimension 1 and its combination with
the antiholomorphic part $\overline{V}_p$ gives the (1,1) vertex operator.
This vertex operator is BRST invariant after integrated over the world sheet.
In other words, its commutator with the BRST operator is a total derivative.
To show this, it is sufficient to show that $[Q, V_p(z)]=\partial O_p(z)$.
$O_p$ is a BRST invariant operator of conformal dimension zero. Now the
holomorphic part of the BRST operator is $Q=\oint dZC(Z)(T(Z)+1/2 T_g(Z))$,
where the super contour integral is defined as $\oint dZ=1/(2\pi i)\oint dz
\int d\theta^-d\theta^+$. $Q$ can be divided into three parts:
\eqn\brst{\eqalign{&Q=Q_1+Q_2+Q_3\cr
&Q_1={1\over 2\pi i}\oint dz(cT_t(z)-cb\partial c(z))\cr
&Q_2={1\over 2\pi i}\oint dz(\gamma^+G^-(z)+\gamma^-G^+(z)+{\tilde{c}\over 2}
J(z))\cr
&Q_3={1\over 2\pi i}\oint dz\left(-{b\over 2}\gamma^+\gamma^-(z)+{\tilde{b}
\over
2}(\gamma^-\partial\gamma^+-\gamma^+\partial\gamma^-)(z)+{\tilde{c}\over
2}(\gamma^-\beta^+-\gamma^+\beta^-)(z)\right),}}
where $T_t(z)$ is the total stress tensor excluding $(b,c)$, $G^{\pm}$ are
the supercurrents for the matter system. $Q_1$ is just like the BRST operator
in a bosonic string when all other ghosts are included in the matter system.
To calculate $[Q, V_p]$, note that the OPE of $V_p$ with the integrand in
$Q_3$ is non-singular, so the nontrivial terms come from its commutators with
$Q_1$ and $Q_2$. Now $[Q_1, V_p]=\partial(cV_p)$ as is expected, and
$[Q_2, V_p]=(1/2)\partial(\gamma^+V^-_p+\gamma^-V^+_p)$. We find
\eqn\inv{O_p=cV_p+{1\over 2}(\gamma^+V^-_p+\gamma^-V^+_p).}
$O_p$ is BRST invariant, as can be checked directly or by the consistency
of the formula $[Q, V_p]=\partial O_p$. Nevertheless, $O_p$ can not be
a BRST commutator, for otherwise $V_p$ would be a total derivative up to a BRST
commutator. We note that our result
is different from that in \dis. To show that $O_p$ is BRST invariant, one needs
the fact that $p\bar{p}=0$ and the U(1) charge of $O_p$ is zero. If one
formally takes $\bar{p}=0$ while $p$ not be zero, $O_p$ is still BRST
invariant. Next take $p\rightarrow 0$ and extract the first order terms in
$O_p$. We find the following BRST invariant operators at $p=0$ \dis\
\eqn\more{O_{\bar{i}}=c\partial\bar{x}^i_L-{1\over 2}\gamma^+\overline{\psi}
^i_L.}
Exchanging $p$ and $\bar{p}$, we also find
\eqn\mor{O_i=c\partial x^i_L+{1\over 2}\gamma^-\psi^i_L.}
These operators are nothing but representatives of  the constant deformations
of the metric and antisymmetric tensor field when combined with their
antiholomorphic counterparts.

As we learned in the N=1 strings, the natural operators to use in the
correlation functions are operators in the $-1$ picture for the NS operators
and the ones in the $-1/2$ picture for the Ramond operators \ref\fms{
D. Friedan, E. Martinec and S. Shenker, Nucl. Phys. B271 (1986) 93.} \ref\vv{
E. Verlinde and H. Verlinde, in Superstrings '88, eds. M. Green et al.
Singapore: World Scientific 1989.}. In the un-twisted
N=2 string, operators can be naturally considered as in the NS sector.
Since there are two pairs of $(\beta, \gamma)$, the natural picture of
operators is then the $(-1,-1)$ picture. In this picture, operators carry
$\gamma^+$ ghost number $-1$ and $\gamma^-$ number $-1$. To construct these
operators, it is necessary to bosonize $(\beta^{\mp}, \gamma^\pm)$.
As in \fms\ we introduce two scalars $\phi_\pm$ and two pairs of
fermionic $(\eta^\mp, \xi^\pm)$. The bosonization relations are
$$\eqalign{&\gamma^+=e^{\phi_+}\eta^+, \quad \beta^-=e^{-\phi_+}\partial
\xi^- \cr
&\gamma^-=e^{\phi_-}\eta^-, \quad \beta^+=e^{-\phi_-}\partial\xi^+.}
$$
A simple conformal dimension zero operator in the $(-1,-1)$ picture is
$$2ce^{-\phi_+}e^{-\phi_-}\tilde{V_p}$$
when $p$ is on-shell.
This operator is BRST invariant and we denote it by $O^{(-1)}_p$.
Consider operators $[Q, \xi^+O^{(-1)}_p]=-O^+_p$ and $[Q, \xi^-O^{(-1)}_p]
=O^-_p$. These operators are certainly BRST invariant. They are nontrivial
however in the Hilbert space containing no zero modes of $\xi^\pm$. Indeed
there are relations $O^\pm_p=c\hbox{exp}(-\phi_\pm)V^\pm_p$. One further
constructs a (0,0) picture operator from $O^\pm_p$ and finds
\eqn\comm{O_p=[Q,\xi^+O^-_p+\xi^-O^+_p].}
So the (0,0) picture operator $O_p$ is related to the $(-1,-1)$ picture
operator $O^{(-1)}_p$ in much the same way as a 0 picture operator
related to a $-1$ picture operator in the N=1 strings. BRST commutators must
be taken twice because of two pairs of bosonic ghosts. $[Q, \xi^+O^-
_p]$ and $[Q, \xi^-O^+_p]$ are two independent BRST invariant operators.
But we shall see later that only their combination appears in
correlation functions.

The ground ring operators considered in \give\ are not BRST invariant operators
and therefore unnatural. One instead considers OPE's of operators
$O^{(-1)}_p$. The product of two such operators generates an operator in
the $(-2, -2)$ picture. When $p_1+p_2$ is on-shell, the product $O^{(-1)}
_{p_1}(z)O^{(-1)}_{p_2}(0)$ has a well-defined limit as $z\rightarrow 0$.
The result is a BRST invariant operator $O^{(-2)}_{p_1+p_2}$ in the
$(-2,-2)$ picture. We shall not bother to write it down. We are interested
in its corresponding operator in the $(-1,-1)$ picture. Indeed it can be
shown that
$$[Q,\xi^-[Q,\xi^+O^{(-2)}_{p_1+p_2}]+\xi^+[Q,\xi^-O^{(-2)}_{p_1+p_2}]]
={1\over 2}(p_1\bar{p}_2-\bar{p}_1p_2)\partial c O^{-1}_{p_1+p_2}.$$
Certainly $\partial cQ^{(-1)}_p$ is BRST invariant. It can be thought of
as the conjugate of $O^{(-1)}_{-p}$ and therefore is not a BRST exact operator.
We would like to conjecture that the only BRST cohomology states in the
$(-1,-1)$
picture are these two sets of operators. If this conjecture is true, one
easily sees that when $p_1+p_2$ is off-shell, the product $O^{(-1)}_{p_1}
(z)Q^{(-1)}_{p_2}(0)$ is BRST exact and therefore can be put zero in the
BRST cohomology. We conclude that
\eqn\ope{O^{(-1)}_{p_1}O^{(-1)}_{p_2}\sim \cases{{1\over 2}
c_{p_1,p_2}\partial cO^{(-1)}_{p_1+p_2} & \qquad \hbox{$p_1+p_2$ on-shell}\cr
0    &\qquad\hbox{$p_1+p_2$ off-shell},}}
where $c_{p_1,p_2}=p_1\bar{p}_2-\bar{p}_1p_2$. We shall see in the next section
that the above structure is similar to the (left) gauge algebra
generated by currents $V_p$ in the case of torus compactification.

\newsec{Gauge Symmetries in Background $T^{2,2}$}

Since the BRST commutator of $V_p$ is a total derivative, one attempts to
use these currents to construct conserved charges. The problem in the
usual flat background $R^{2,2}$ is that the left momentum must match the
right momentum, thus the conserved charges are not truly conserved charges.
As we shall see later, if one uses charges constructed from currents
$V_p$ and formally derives Ward identities associated to these charges,
the nontrivial three point tree amplitudes will violate such Ward identities.
The authors of \give\ then suggest one instead consider compactified
space-time $T^{2,2}$. Now the matching condition for the left momentum
$p_L$ and the right momentum $p_R$ is that $(p_L, p_R)$ lies on a self-dual
even lattice $\Gamma^{4,4}$. Now there are many on-shell $p$'s such that
$(p,0)$ are on this lattice. The charges constructed as
$$Q_p={1\over 2\pi i}\oint V_pdz$$
are conserved, BRST invariant.

Next we make an important conjecture. The only nontrivial BRST cohomology
operators in the $(-1,-1)$ picture are products of $O^{-1}_{p_L}$ and
$\partial c O^{(-1)}_{p_L}$ with $\overline{O}^{(-1)}_{p_R}$ and
$\overline{\partial}\bar{c}\overline{O}^{(-1)}_{p_R}$ for $(p_L,p_R)
\in\Gamma^{4,4}$. This conjecture implies a similar statement for nontrivial
BRST cohomology operators in the $(0,0)$ picture, with exceptions
of operators at zero momentum. Given this conjecture,
the Lie algebra generated by $Q_p$ can be calculated. When $p_1+p_2$ is not
on-shell, we shall argue that the commutator $[Q_{p_1},Q_{p_2}]$ is
zero up to a BRST exact operator. The argument goes as follows.
First, since $(p_1,0)$ and $(p_2,0)$ are both on the lattice $\Gamma^{4,4}$,
$p_1\bar{p}_2+\bar{p}_1p_2\in 2Z$. This implies that $[Q_{p_1},Q_{p_2}]$
can be expressed as a contour integral. Let the integrand be $V'_{p_1+p_2}$.
Since the commutator is BRST invariant, then the commutator of $V'_{p_1+p_2}$
with the BRST operator is a total derivative $\partial O'_{p_1+p_2}$.
$O'_{p_1+p_2}$ is BRST invariant by a standard argument. This operator in
turn must be a BRST exact operator, as implied by our conjecture.
Using this fact and the identity $[Q,V'_{p_1+p_2}]=\partial O'_{p_1+p_2}$, one
easily sees that $V'_{p_1+p_2}$ is a total derivative up to a BRST exact
operator. This results in what we wanted to prove. When $p_1+p_2$ is
on-shell, we calculate $[Q_{p_1}, Q_{p_2}]=c_{p_1,p_2}Q_{p_1+p_2}$. This
relation is first derived in \give. We conclude that
\eqn\algebra{[Q_{p_1},Q_{p_2}]=\cases{c_{p_1,p_2}Q_{p_1+p_2} &\qquad
\hbox{$p_1+p_2$ on-shell}\cr
0 & \qquad \hbox{$p_1+p_2$ off-shell}}.}
This relation is similar to the OPE in \ope. We should point out that
because of the second equality in \algebra, the Lie algebra we just
found is different from the one suggested in \give. Similarly conserved
charges $Q_i$ and $Q_{\bar{i}}$ can be constructed from currents
$\partial x^i_L$, $\partial\bar{x}^i_L$.

It is necessary to check that \algebra\ indeed defines a Lie algebra,
namely the Jacobi identity holds. Consider $[[Q_{p_1},Q_{p_2}],Q_{p_3}]+
[[Q_{p_3},Q_{p_1}], Q_{p_2}]+[[Q_{p_2},Q_{p_3}],Q_{p_1}]$. If $p_1+p_2+p_3$
is not on-shell, every term in this expression is zero. So the Jacobi
identity hold in this case. Suppose $p_1+p_2+p_3$ is on-shell. There are
three possibilities. 1. When all three $p_i+p_j$, $i\ne j$ are off-shell,
each term is zero in the above expression. 2. Two of $p_i+p_j$ are
on-shell, then the third is on-shell too, for $p_1+p_2+p_3$ is on-shell.
The Jacobi identity can be checked readily, using the first equality in
\algebra. 3. Only one of $p_i+p_j$ is on-shell. Let it be $p_1+p_2$.
Now in the Jacobi identity the second term and the third term are zero.
The first term is $c_{p_1,p_2}c_{p_1+p_2,p_3}Q_{p_1+p_2+p_3}$.
$p_2$ is independent of $p_1$, otherwise $p_1+p_3$ must be on-shell.
By using rotations of $SO(2,2)$ we can always put $p_1=(1,1)$ (in the
notation $p=(p^1,p^2)$). $p_1+p_2$ is on-shell so $p_2\sim (e^{i\phi}, e^{\pm i
\phi})$. When $p_2\sim (e^{i\phi},e^{i\phi})$, $c_{p_1,
p_2}=0$. So the first term in the Jacobi identity is zero too.
When $p_2\sim (e^{i\phi},e^{-i\phi})$, $c_{p_1,p_2}$
is not zero. Assume $p_3\sim (\hbox{exp}(i\psi),\hbox{exp}(i\theta))$.
Now $p_1+p_2\sim (\hbox{exp}(i\phi'),\hbox{exp}(-i\phi'))$, $\phi'$ is
another phase, different from $\phi$. Since
$p_1+p_2+p_3$ is on-shell, there are only two possible cases for $p_3$.
The first is $\psi=\theta+2\phi'$ mod $2\pi$. The second case is
$\psi=-\theta$ mod $2\pi$. The latter is ruled out by the fact that
$p_1+p_3$ is off-shell. For the first case, it is readily seen that
$c_{p_1+p_2,p_3}=0$. So the Jacobi identity holds again. This check of
validity of the Jacobi identity supports the second equality in \algebra,
which is a consequence of the conjecture made before.

Now consider the action of the algebra on BRST invariant operators
$O^{(-1)}_p$. Again by our conjecture one draw the conclusion that
$[Q_p, O^{(-1)}_q]$ is a BRST exact operator if $p+q$ is off-shell.
When $p+q$ is on-shell, $[Q_p,O^{(-1)}_q]=(1/2)c_{p,q}O^{(-1)}_{p+q}$.
So operators $O^{(-1)}_p$ form a module of the algebra \algebra.
It can be checked that the representation is consistent with the algebra.

\newsec{Amplitudes and Ward identities}

As shown by Ooguri and Vafa, except for some non-trivial three point tree
amplitudes, all other tree amplitudes should be zero. They explicitly
calculated four point tree amplitudes, and proved that due to an identical
vanishing  kinematical factor all four point three amplitudes are zero.
This is consistent with the low energy action which must be exact to all
order in perturbations, as argued by these two authors. This argument
for vanishing of higher amplitudes is rather indirect and a direct argument
is desired. We shall show in this section that on a torus compactification
$T^{2,2}$, the Ward identities associated to those conserved charges
discussed in the last section actually can be used to derive at least
a part result about amplitudes of arbitrary loops. Non-trivial three point
tree amplitudes are checked to satisfy Ward identities. A large class
of four point amplitudes of arbitrary loops are shown to be vanishing.
We have not tried to draw consequences of the Ward identities for
higher point amplitudes. It appears that these identities alone do
not determine all amplitudes, as being obvious from our discussion
about three point amplitudes. To get a similar
result for strings moving in background $R^{2,2}$, one simply takes a limit
of certain torus compactification.

\noindent $\underline{\hbox{\it Amplitudes}}$

We begin with a definition of amplitudes of loop $g$. First we shall
work on a Riemann surface of genus $g$. Consider a correlation function
of $n$ BRST invariant operators ${\cal O}_i$, $i=1,\dots , n$.
\eqn\corr{\langle {\cal O}_1{\cal O}_2\dots {\cal O}_n\rangle.}
Let the $c$ ghost number of ${\cal O}_i$ be $c_i$, the $\gamma^\pm$ ghost
number $\gamma^\pm_i$ and the $\tilde{c}$ ghost number $\tilde{c}_i$. Since
the vacuum associated to a Riemann surface of genus $g$ has respective
ghost numbers $6g-6$, $4-4g$ and $2g-2$, the total ghost numbers in the
above correlation are
$$\eqalign{2s&=6g-6+\sum_{i=1}^n c_i\cr
-2s^\pm&=4-4g+\sum_{i=1}^n\gamma^\pm_i\cr
2\tilde{s}&=2g-2+\sum_{i=1}^n \tilde{c}_i}.$$
Thus the correlation in \corr\ is not well-defined unless these net ghost
numbers be balanced. We then insert a number of surface integrals $\int
\eta_ab$ and their anti-holomorphic counterparts to cancel the $c$ ghost
number. Here $\eta_a$ are dimension $(-1,1)$ Beltrami differentials. Similarly
to cancel the $\gamma^\pm$ ghost numbers we insert a number of $
\delta(\int \eta^\pm_{a^\pm}\beta^\mp)$ (and anti-holomorphic counterparts).
$\eta^\pm$ are dimension $(-1/2, 1)$ Beltrami differentials. Finally
a number of integrals $\int\tilde{\eta}_{\tilde{a}}\tilde{b}$ (and
anti-holomorphic counterparts) are to be inserted to cancel the $\tilde{c}$
ghost number, $\tilde{\eta}_{\tilde{a}}$ are dimension $(0,1)$ Beltrami
differentials. Recall that we are dealing with a $N=2$ string, the Riemann
surface is indeed a $N=2$ super Riemann surface \ref\cohn{J.D. Cohn,
Nucl. Phys. B284 (1987) 349.}.
There is a super moduli space of super Riemann surfaces of genus $g$.
Beltrami differentials $\eta$, $\eta^\pm$ and $\tilde{\eta}$ indeed
represent tangent vectors at a point on the super moduli space. Correlation
\corr\ then defines a super form on such super moduli space via
\eqn\form{\eqalign{ F_n(\eta_a,\eta^\pm_{a^\pm},\tilde{\eta}_{\tilde{a}},\dots)
&=\langle {\cal O}_1\dots {\cal O}_n\prod_{a=1}^s\int\eta_ab\prod_{a^+=1}^{s^+}
\delta\left(\int\eta^+_{a^+}\beta^-\right)\cr
&\prod_{a^-=1}^{s^-}\delta\left(\int\eta^-_{a^-}\beta^+\right)\prod_{\tilde{a}
=1}^{\tilde{s}}\int\tilde{\eta}_{\tilde{a}}\tilde{b}\dots\rangle ,}}
where $s$, $s^\pm$ and $\tilde{s}$ are assumed be non-negative integers and
the same number of insertions of anti-holomorphic partners are neglected.
$F_n$ is a closed form on the super moduli space, as can be shown along the
line in \ref\nelson{H.S. La and P. Nelson, Nucl. Phys. B332 (1990) 83.}.
Since all BRST invariant operators in the present theory of interest are
in the NS sector, the dimension of odd moduli associated to
each world sheet gravitino is $4g-4+2n$. The dimension of moduli associated
to flat $U(1)$ connections is $2g-2+2n$. The condition for $F_n$ to be a
top form
on the super moduli space is $\sum c_i=\sum \tilde{c}_i=
-\sum \gamma^\pm_i=2n$. The natural candidates of operators that
make this condition fulfilled are ${\cal O}_i
=\tilde{c}\bar{\tilde{c}}O^{(-1)}_{p_i}\overline{O}^{(-1)}_{p'_i}$,
where $(p_i,p'_i)\in\Gamma^{4,4}$. Other operators are not annihilated
by $b_0-\bar{b}_0$ \foot{Operator $\tilde{c}\bar{\tilde{c}}(\partial c
+\overline{\partial}\bar{c})O^{(-1)}_p\overline{O}^{(-1)}_{p'}$ is
annihilated by $b_0-\bar{b}_0$ but its $c$ ghost number is not appropriate
here. It may play a role elsewhere.}, a necessary condition for the amplitude
to be well-defined \nelson.  When $g> 1$, we can choose
n of $\eta_a$ to be associated to the moduli of punctures. The same can
be done for $\eta^\pm$ and $\tilde{\eta}_{\tilde{a}}$. Indeed those
n of $\tilde{\eta}_{\tilde{a}}$ are chosen such that each of
$\int\tilde{\eta}_{\tilde{a}}\tilde{b}$ becomes $\oint\tilde{b}$
with a contour surrounding only one puncture, and this contour integral
together with its anti-holomorphic partner turns the corresponding
${\cal O}_i$ into $O^{(-1)}_{p_i}\overline{O}
^{(-1)}_{p'_i}$. Since these operators carry zero $U(1)$ charge, the
integration of the $U(1)$ puncture moduli results in a constant. Hereafter
we shall ignore these moduli.

The integration of odd puncture moduli deserves a detailed discussion.
Let $m_a^\pm$ be the odd moduli associated to the Beltrami differential
$\eta_a^\pm$. In a path integral formulation, integration over $m_a^\pm$ will
bring down from the action a factor $\int \eta^\pm_a G^\mp$. If $m_i^\pm$
is an odd moduli of puncture $i$, this factor can be written as
$\oint_{C_i} G^\mp$, the contour $C_i$ surrounds puncture $i$ only.
Together with the insertion of delta functions, we obtain a factor
$\oint_{C_i} G^-\oint_{C_i}G^+\delta(\oint_{C_i}\beta^+)\delta(\oint_{C_i}
\beta^-)$. Now use the relation $\oint_{C_i}G^\pm = [Q,\oint_{C_i}\beta^\pm
]$, the factor can be written as $[Q, H(\oint_{C_i}\beta^+)][Q, H(\oint_{C_i}
\beta^-]$, where $H(\cdot)$ is the Heaviside step function. Going further,
we encounter a ambiguity as how to arrange two contours in the two contour
integrals $\oint_{C_i}\beta^+$ and $\oint_{C_i}\beta^-$. For the time being
we assume the contour in the first contour integral surrounds the one in
the second contour integral. Then by the standard argument \vv\ the picture
changing operator turns $O^{(-1)}_{p_i}$ in the operator $O^{(-1)}_{p_i}
\overline{O}^{(-1)}_{p_i'}$ into $[Q,\xi^+[Q,\xi^- O^{(-1)}_{p_i}]]$.
Leting $Q^\pm$ denote operators $[Q, H(\int_{C_i}\beta^\pm)]$, the above
picture-changed operator is just $Q^+Q^-O^{(-1)}_{p_i}$.
Now note that the moduli $m_i^\pm$ entering path integral together
with $Q^\pm$ must be in either the order $(m_i^-Q^+)(m_i^+Q^-)$ or the order
$(m_i^+Q^-)(m_i^-Q^+)$. These two must be on the same footing, so the
total insertion is then $m_i^-m_i^+[Q^+,Q^-]$. Integrating over $m_i^\pm$
we are left with picture changing operator $[Q^+, Q^-]$ which turns
$O^{(-1)}_{p_i}$ into
$$\left[Q, \xi^+[Q, \xi^-O^{(-1)}_{p_i}]-\xi^-[Q,\xi^+O^{(-1)}_{p_i}]
\right].$$
This is just the operator $O_{p_i}$, see \comm. The integration of
even puncture moduli further changes $O_{p_i}$ into $V_{p_i}$.

A simple application of picture changing operation discussed above
to three point tree amplitudes tells us that these amplitudes can be
simply calculated as correlation function $\langle O^{(-1)}_{p_1}
O^{(-1)}_{p_2}O_{p_3}\rangle$ together with its anti-holomorphic
partner. The result is $c_{p_1,p_2}c_{p'_1,p'_2}$. We obtain the
same formula give in \ov\ when the left momenta are the same as the
right momenta.

\noindent $\underline{\hbox{\it Ward Identities}}$

We now employ the conserved charges described in the last section to
derive Ward identities. The derivation is very simple, for the current
$V_p$ does not involve ghosts and has no antiholomorphic component.
The same Ward identities can be derived for $\overline{Q}_p$ constructed
from $\overline{V}_p$. Insert the charge $\oint_C V_p$ into the correlation
function in \form, $C$ is a contractible contour on the Riemann surface.
To get a sensible result, the total momentum is conserved, $p+\sum_i p_i
=0$. Homologically deform the contour $C$ into contours $C_i$ surround
every puncture. There is no problem for the contour to pass through
points in the integrations $\int \eta b$, $\int \eta^\pm\beta^\mp$, since
the OPE's of $V_p$ with ghosts are not singular. Now $Q_p$ acts on every
$O^{(-1)}_{p_i}$ and gives either zero when $p+p_i$ not null or $(1/2)
c_{p,p_i}
O^{(-1)}_{p+p_i}$ when $p+p_i$ null. On the other hand, the result after
insertion of the charge must be zero, for initially $C$ is shrinkable.
Let $S_p$ be the set of all $p_i$ with the property that $p+p_i$ is
on-shell, we obtain a simple Ward identity
\eqn\ward{\sum_{p_i\in S_p} c_{p,p_i}A_{p_1,\dots, p+p_i,\dots, p_n,\dots}=
0,}
where we used $A_{p_1,\dots p_n,\dots}$ to denote the integrated amplitude
and the dots after $p_n$ denote the right momenta. Note that the Ward identity
\ward\ involves amplitudes of the same loop with the same right momenta.
A similar Ward identity can be derived for insertion of $\overline{Q}_p$.
Remember that the momentum $p$ in \ward\ is not arbitrary, $(p,0)$ must
be a vector
on lattice $\Gamma^{4,4}$. Ward identities associated to $Q_i$ and
$Q_{\bar{i}}$ are just the conservation law of momentum.

We should check this Ward identity for three point tree amplitudes. Of course
we are interested in the case when $S_p$ is non-empty. Say $p_1\in S_p$.
Since $p_1+p_2+p_3+p=0$, $p+p_1+p_2$ is on-shell. If $p_1+p_2$ is on-shell,
then $p+p_2$ is on-shell and so is $p+p_3$. So $S_p$ contains all three
$p_i$. As we have seen all three amplitudes involved in the Ward identity
have the same right momenta and so we can forget about the contribution
from the right sector to the amplitudes. It is then easy to show that
$c_{p,p_1}c_{p+p_1,p_2}+c_{p,p_2}c_{p_1,p+p_2}+c_{p,p_3}c_{p_1,p_2}=0$.
Thus the Ward identity holds in this case. The other case is when $p_1+p_2$
is not null and so neither $p+p_2$ and $p+p_3$. $S_p$ contains only
$p_1$. The Ward identity reads $c_{p,p_1}c_{p+p_1,p_2}=0$. This situation
is exact the same as in case 3 in our proof of the Jacobi identity
in the last section, provided we identify $p$, $p_1$, $p_2$ here with
$p_1$, $p_2$, $p_3$ there. So we have proved that non-trivial three point
tree amplitudes satisfy the Ward identity. Were there a similar Ward identity
for three point tree amplitudes in the background $R^{2,2}$, all amplitudes
involved would contain matched left and right momenta. In this case an
amplitude takes of form $c_{p_1,p_2}^2$. It is straightforward to show that
the supposed Ward identity is violated.

More can be said about three point amplitudes of arbitrary loops. The
identity \ward\ may even be regarded as an identity for non-perturbative
amplitudes, since its structure does not depend on the loops. For simplicity
let us consider a lattice $\Gamma^{4,4}=\Gamma^{2,2}\oplus \Gamma^{2,2}$.
Both $\Gamma^{2,2}$ are even self-dual, and the first is of the left
momenta and the second is of the right momenta. So we can work on the
left momenta without paying attention to the right ones. Before give a
general result for the three point amplitudes, let us state a fact
concerning the lattice $\Gamma^{2,2}$ which is proved in the appendix.
Given any null vector $p$ on $\Gamma^{2,2}$, there are two different
maximal null sublattices of $\Gamma^{2,2}$ containing $p$. Each lattice is
two dimensional. On one sublattice, every momentum $p'$ satisfies $c_{p,p'}
=0$. Denote this lattice by $\Gamma(p)$. On the other sublattice, any $p'$
independent of $p$ satisfies $c_{p,p'}\ne 0$. Denote this lattice by
$\Gamma'(p)$.

Now consider a three point amplitude of any loops $A_{p_1,p_2,p_3,\dots}$,
if we like, we can even assume it is a nonperturbative amplitude. Since
$p_1\cdot p_2=0$, $p_2$ must be on one of the maximal null lattices
containing $p_1$. If $c_{p_1,p_2}=0$ and $p_2$ is independent of $p_1$,
we choose a $p$ on the null lattice $\Gamma'(p_1)$. Apparently a $p$
can be chosen such that $p\cdot p_2\ne 0$, otherwise $p_2$ would be on this
lattice. So $p+p_2$ is not null and neither is $p+p_3$. Decomposing
$p_1=p+p_1'$ and using the Ward identity, we find $A_{p_1,p_2,p_3,\dots}=
0$. The other possibility is that $p_2$ is on $\Gamma'(p_1)$. In this case
it takes more effort to obtain a concrete result. Let us start with
an amplitude $A_{np,mp,-(n+m)p}$, $n, m$ are integers. Hereafter we shall
omit dots representing the right momenta in the subscript. Choose a $p'$
on $\Gamma'(p)$, the Ward identity reads
\eqn\any{c_{p',p}\left(nA_{np,mp,-(n+m)p}+mA_{np-p',mp+p',-(n+m)p}
-(n+m)A_{np-p',mp,-(n+m)p+p'}\right)=0.}
Taking $n=0$ we obtain $mA_{-p',mp+p',-mp}-mA_{-p',mp,-mp+p'}$, if $c_{p'
p}\ne 0$. Note that
a three point amplitude is a function of the first two momenta. This Ward
identity suggests that if the first momentum is $-p'$, adding a multiple
of $p'$ to the second momentum does not change the amplitude. This line of
argument shows that if $(e_1,e_2)$ are generators of a null lattice with
$c_{e_1,e_2}\ne 0$, then $A_{e_1,me_1+ne_2, -(m+1)e_1-ne_2}=A_{e_1,ne_2,-
e_1-ne_2}$. Note that in this equality $n$ can not be zero. If however
this formula extends to the case $n=0$, we find $A_{e_1,me_1,-(m+1)e_1}=0$,
since the second amplitude involves a zero momentum. This is certainly
true for a tree amplitude. Indeed taking this as an assumption and using
\any\ repeatedly, we find
$$A_{ae_1+be_2,ce_1+de_2, -(a+c)e_1-(b+d)e_2}=(ad-bc)A_{e_1,e_2,-e_1-e_2}.$$
This formula shows that amplitude $A_{e_1,e_2,-e_1-e_2}$ depends only on
the null lattice $\Gamma$ if the orientation of $(e_1,e_2)$ is preserved.
Taking this as a constant proportional to $c_{e_1,e_2}$ which is also
invariant provided the orientation $(e_1,e_2)$ does not change, then
our above result amounts to saying that the amplitude $A_{p_1,p_2,p_3}
=Ac_{p_1,p_2}c_{p'_1,p'_2}$, where $A$ is a constant and we included the
contribution from both sectors. Remember that this result is derived
under the assumption that the amplitude vanishing when $p_1$ and $p_2$
are proportional. This appears not a consequence of Ward
identities. Furthermore, the constant $A$ can not be determined by
use of Ward identities, since these identities are homogeneous. It
is plausible that this constant encodes topological information of
various moduli spaces as well as information of the lattices $\Gamma'
(p_1)$ and $\Gamma'(p'_1)$. When space-time is $R^{2,2}$, it is found
in \ref\it{M. Bonini, E. Gava
and R. Iengo, Mod. Phys. Lett. A6 (1991) 795.}
that the amplitude is proportional to $c^6_{p_1,p_2}$ with a infinite
constant. This was checked and generalized in \ov\ to the space-time
$T^2\times R^2$. Indeed, looking back at \form\ and remembering that
the three odd moduli of the punctured torus for each gravitino must
be attributed to three punctures, one finds that all three operators
are in the $(0,0)$ picture. This implies that the one-loop amplitude
should be proportional to $c^3_{p_1,p_2}c^3_{p'_1,p'_1}$, a contradiction
to the result indicated by the Ward identity. If the infinity can
be regularized in the compactified space-time, the only solution
to this contradiction is that $A=0$ at one loop level. Explicit
calculation is required to confirm this. Similar argument leads to
$A=0$ at any higher loops level \foot{I should thank C. Vafa for a
question leading to this discussion.}.

Finally we apply Ward identities to four point amplitudes. Again we will
see that although these Ward identities imply that many amplitudes are
vanishing, they do not help in some exceptional cases. Naturally we shall
assume that all momenta are not zero, otherwise the amplitude is automatically
vanishing. We consider various cases in the following. First suppose
there are three independent
momenta in $A_{p_1,p_2,p_3,p_4}$. Since $\sum p_i=0$, any three of $(p_i)$
are independent. Consider inner products $p_i\cdot p_j$, $i\ne j$ of these
three momenta. There is at least one nonvanishing $p_i\cdot p_j$. Moreover,
by a suitable choice of these three momenta, there are two nonvanishing
$p_i\cdot p_j$. We assume these three momenta are $p_1$, $p_2$ and $p_3$ and
$p_1\cdot p_2\ne 0$, $p_1\cdot p_3\ne 0$. Now $p_2$ and $p_3$ are not on
the null lattice $\Gamma'(p_1)$. It is possible to choose a $p$ on this
lattice such that $c_{p,p_1}\ne 0$ and $p\cdot p_2\ne 0$, $p\cdot p_2\ne 0$.
Furthermore, if $p_4=-(p_1+p_2+p_3)$ is not on the null lattice, $p$
can be chosen with $p\cdot p_4\ne 0$. So $p+p_i$ $i>1$ are not null. Let
$p_1=p+p'_1$ in \ward\ and it reads $c_{p,p_1}A_{p_i}=0$, namely
$A_{p_i}=0$. If $p_4$ is on the null lattice $\Gamma'(p_1)$, this means
$p_2+p_3$ is on this lattice and particularly $p_2\cdot p_3=0$. Choose
$p=p_4$. Since $p_2$ and $p_3$ are independent of $p_1$, so $c_{p_4,p_1}
\ne 0$. Moreover, $p_4\cdot p_2=-p_1\cdot p_2\ne 0$ and $p_4\cdot p_3
=-p_1\cdot p_3\ne 0$. The Ward identity with $p_1=p+p'_1$ again implies
that $A_{p_i}=0$

Next we consider the case in which there are only two independent momenta
in the amplitude. Let $p_1$, $p_2$ be such two momenta. There are two
cases. 1. If $p_1+p_2$ is
not on-shell, namely $p_1\cdot p_2\ne 0$, we choose a $p\in \Gamma'(p_1)$
with $p\cdot p_2\ne 0$. $p_3$ and $p_4$ are not independent of $p_1$ and
$p_2$. If the former two momenta both contain a nonvanishing component in
$p_2$, then $p+p_3$ and $p+p_4$ are not null. The Ward identity implies
$A_{p_i}=0$. When $p_3$ and $p_4$ both contain a nonvanishing component
in $p_1$, a $p$ on $\Gamma'(p_2)$ can be chosen such that $p+p_i$
$i\ne 2$ are off-shell. Again the Ward identity implies $A_{p_i}=0$.
So the only exceptional case is $A_{p_1,p_2,-p_1,-p_2}$. We can have
an equality $A_{p_1,p_2,-p_1,-p_2}=A_{p_1-p,p_2,-p_1+p,-p_2}$,
if $p\in \Gamma_{p_1}$ and $c_{p,p_1}\ne 0$. Obviously $p$ can be
chosen such that $p_1-p+p_2$ is on-shell. Since $\Gamma'(p_1-p)
=\Gamma'(p_1)$ (see appendix), $p_2$ can not be on this lattice, but
on $\Gamma(p_1-p)$. So the second amplitude in the equality falls into
the other cases we are about to consider. 2. If $p_1+p_2$ is on-shell,
there are two sub-cases. The first sub-case is that $p_2$ is not on
$\Gamma'(p_1)$, again we follow argument in case 1 and show that
the amplitude is zero unless when the amplitude is $A_{p_1,p_2,-p_1,-p_2}$.
The second sub-case is when $p_2$ is on $\Gamma'(p_1)$. The Ward
identity does not help in this case. Note that all $p_i$ is on the
lattice $\Gamma'(p_1)$ in this case. This also includes the case
when $p_i$ are all proportional. If we assume the amplitude is vanishing
when all $p_i$ are proportional, we can again do more in exceptional
cases. We will not do so here.

\newsec{Conclusions}

We have seen in this paper that the study of conserved charges is utmost
important in a string model. Not only it is technically helpful in
calculating amplitudes, also it is linked to deep principles of the
theory. Using those more or less apparent conserved charges, one
already can obtain some concrete results about amplitudes. $N=2$ strings,
just like the $d=2$ string which has been studied intensively recently,
have a rich structure and more is to be unraveled. They serve as an
interesting toy model for understanding deep issues in a realistic
string model. Moreover, as already
pointed out in \ov, they may be an organizing model of lower dimensional
integrable models.

It may be that to get stronger results about amplitudes in $N=2$ strings,
one should study them in more a complicated background in order to
employ more symmetries. One should also interpolate symmetries
in different backgrounds as the authors of \give\ attempted, and this may
be powerful enough to help one to determine all amplitudes. A
possible direction is to discuss the dependence of the gauge algebra on
moduli associated to marginal operators constructed from ``discrete''
operators at zero momentum, and find out the whole background
independent gauge algebra.
Finally, it is straightforward to extend our approach to $N=2$ heterotic
strings \oov.

\noindent {\bf Acknowledgements}

I would like to thank C. Vafa for correspondence, and S. Giddings, K. Li
and C.J. Zhu for conversations. This work was supported by DOE grant
DOE-76ER70023.

\appendix{}{}

It will be shown in this appendix that for a null $p$ on an even self-dual
lattice $\Gamma^{2,2}$, there are two maximal null sublattices containing
$p$. These null lattices are two dimensional. On one null lattice, any
$p'$ satisfies $c_{p',p}=0$. This null lattice is denoted as $\Gamma(p)$.
On the other null lattice, any $p'$ independent of $p$ satisfies $c_{p',p}
\ne 0$. This null lattice is denoted as $\Gamma'(p)$. To see that there
are only two maximal null sublattices containing $p$, consider equation
$p\cdot k=0$. Solutions $k$ to this equation lie on a three dimensional
space. These solutions module $\alpha p$, $\alpha$ is real number, lie
on a two dimensional space $R^{1,1}$ with signature $(1,1)$. We are interested
in null spaces spanned by $p$ and solutions to $p\cdot k=0$. Apparently
the number of these spaces are just the number of independent null
vectors in $R^{1,1}$. There are only two independent null vectors in
such a space. We then conclude that there are at most two maximal
null lattices containing $p$. Next we should show that these null lattices
indeed exist. We show this by construction. Recall that any even
self-dual lattice $\Gamma^{2,2}$ by a Lorentzian rotation is equivalent
to the lattice generated by $(e_1,e_2, e_3, e_4)$, where all $e_i$ are
null, and among inner products $e_i\cdot e_j$ $i\ne j$ only $e_1\cdot e_2
=e_3\cdot e_4=1$ are non-vanishing. Choose the convention
$c_{e_1,e_3}=c_{e_2,e_4}=0$ and others zero. Let $p=m_1e_1+m_2e_2+m_3e_3
+m_4e_4$, then the vector $p'=-m_4e_1-m_3e_2+m_2e_3+m_1e_4$ is null
and orthogonal to $p$. It is also independent of $p$ and $c_{p',p}
=0$. Thus there is a null lattice $\Gamma(p)$ on which $p$, $p'$ as
are two independent vectors. Any vector on this lattice
decomposes into components in $p$ and $p'$ with rational coefficients.
Next consider $p'=-m_3e_1-m_4e_2+m_1e_3+m_2e_4$. This vector is
null, orthogonal to and independent of $p$. Moreover $c_{p',p}\ne 0$.
This shows that $\Gamma'(p)$ exists and any vector on it decomposes
into components in $p$ and $p'$ with rational coefficients. By this
construction proof, it can be seen that for any $p'\in \Gamma'(p)$,
$\Gamma'(p')=\Gamma'(p)$. So two different such null lattices do not
intersect.

\vfill\eject
\listrefs\end